\documentclass[aps,pra,twocolumn,superscriptaddress,groupedaddress, showkeys]{revtex4}  

\usepackage{graphicx}  
\usepackage{dcolumn}   
\usepackage{bm}        
\usepackage{amssymb}   
\usepackage{color}
\usepackage{epstopdf}
\usepackage{amsmath}

\begin{document}

\title{Temperature-dependent optical spectra of single-crystal (CH$_3$NH$_3$)PbBr$_3$  cleaved in ultrahigh vacuum}

\author{Daniel~Niesner} 
\email{daniel.niesner@fau.de}
\author{Oskar Schuster}
\author{Max Wilhelm}
\affiliation{Lehrstuhl f\"ur Festk\"orperphysik, Friedrich-Alexander-Universit\"at Erlangen-N\"urnberg (FAU), Staudtstr.~7, 91058~Erlangen, Germany}

\author{Ievgen Levchuk} 
\author{Andres Osvet}
\author{Shreetu Shrestha}
\author{Miroslaw Batentschuk}
\affiliation{Institute of Materials for Electronics and Energy Technology~(I-MEET), Department of Materials Science and Engineering, Friedrich-Alexander-Universit\"at Erlangen-N\"urnberg (FAU), Martensstrasse~7, 91058~Erlangen, Germany}

\author{Christoph Brabec}
\affiliation{Institute of Materials for Electronics and Energy Technology~(I-MEET), Department of Materials Science and Engineering, Friedrich-Alexander-Universit\"at Erlangen-N\"urnberg (FAU), Martensstrasse~7, 91058~Erlangen, Germany}
\affiliation{Bavarian Center for Applied Energy Research (ZAE Bayern), Haberstrasse 2a, 91058 Erlangen, Germany}

\author{Thomas Fauster}
\affiliation{Lehrstuhl f\"ur Festk\"orperphysik, Friedrich-Alexander-Universit\"at Erlangen-N\"urnberg (FAU), Staudtstr.~7, 91058~Erlangen, Germany}

\keywords{Organic-inorganic perovskite, hot carrier, photoluminescence, CH3NH3PbBr3, electron-phonon coupling}

\date{\today}

\begin{abstract}


We measure temperature-dependent one-photon and two-photon induced photo\-luminescence from (CH$_3$NH$_3$)PbBr$_3$ single crystals cleaved in ultrahigh vacuum. An approach is presented to extract absorption spectra from a comparison of both measurements. Cleaved crystals exhibit broad photoluminescence spectra. 
We identify the direct optical band gap of 2.31~eV. Below 200~K the band gap increases with temperature, and it decreases at elevated temperature, as described by the Bose-Einstein model. An excitonic transition is found 22~meV below the band gap at temperatures $<200$~K. Defect emission occurs at photon energies $<2.16$~eV.   
In addition, we observe a transition at 2.25~eV (2.22~eV) in the orthorhombic (tetragonal and cubic) phase. Below 200~K, the associated exciton binding energy is also $22$~meV, and the transition redshifts at higher temperature. The binding energy of the exciton related to the direct band gap, in contrast, decreases in the cubic phase. 
High-energy emission from free carriers is observed with higher intensity than reported in earlier studies. It disappears after exposing the crystals to air.


\end{abstract}

\maketitle

\section{Introduction}

Organic-inorganic perovskite semiconductors (OIPS) have opened a whole new field in opto\-electronics~\cite{Yang2015, saliba2016, stranks2015, dou2014, Yakunin2015, ha2015, zhu2015}. A comprehensive understanding of the underlying photo\-physics is still under development. Therefore, detailed knowledge is required of the band structure of OIPS~\cite{even2013, brivio2014, quarti2014, kawai2015, zheng2015}, modifications by local disorder~\cite{mashiyama1998, etienne2016, azarhoosh2016, even2016, beecher2016, yaffe2016}, excitonic effects~\cite{stranks2014, even2014a, Menendez2015, miyata2015, galkowski2016, demchenko2016}, polaronic screening~\cite{bokdam2015, yu2016, Frost2016, zhu2016}, and their interplay. Many studies that tackle these questions rely on the interpretation of (time-resolved) optical spectroscopy. Experimental data and their interpretation, however, show strong variations. Spectroscopic results obtained from thin films depend on growth technique~\cite{park2015,grancini2015,xing2015, wu2015, dai2016}, grain size~\cite{grancini2015, li2016a}, and environmental conditions~\cite{fang2016, muller2015}. 

For example, exciton binding energies between 15 and 84~meV~\cite{Tanaka2003, galkowski2016, yang2015a, yang2015c, green2015, zheng2015a, tilchin2016} have been reported for (CH$_3$NH$_3$)PbBr$_3$.  These variations can be expected to be reduced by studying single crystals~\cite{shi2015, dong2015}, opening the opportunity to approach the intrinsic photo\-physics of OIPS. However,  band gaps of (CH$_3$NH$_3$)PbBr$_3$ single crystals determined from reflection measurements vary from 2.22 to 2.35~eV~\cite{murali2016, yang2015c, kunugita2015, yang2015a, wu2016}. 
Exposure of crystal surfaces to air results in fast hydration, increasing the room-temperature band gap from 2.22~eV to 2.27~eV~\cite{murali2016}.  Recent experimental findings made on lead-bromide single-crystals also include the observation of photo\-luminescence (PL) from high-energy carriers~\cite{zhu2016} with the potential to enhance solar cell performance, possibly beyond the Shockley-Queisser limit~\cite{ross1982}. High-energy carriers have not been apparent to the same degree in earlier studies on thin films. 
Along another line of research, studies on (CH$_3$NH$_3$)PbI$_3$ thin films reveal the coexistence of a direct and an indirect optical band gap with measured energetic spacings of 47~meV~\cite{hutter2016} and 60~meV~\cite{wang2016}. A slightly indirect band gap has been proposed as one of the origins of the low carrier recombination rate found in OIPS~\cite{zheng2015, etienne2016, azarhoosh2016}. 
A possible direct-indirect character of the band gap of related (CH$_3$NH$_3$)PbBr$_3$ remains to be investigated. Photoluminescence spectra of (CH$_3$NH$_3$)PbBr$_3$ showed a double-peak structure~\cite{fang2015, wu2016, murali2016}, and further studies are needed to identify its origin. 

\begin{figure}
\center\includegraphics[width=0.43\columnwidth]{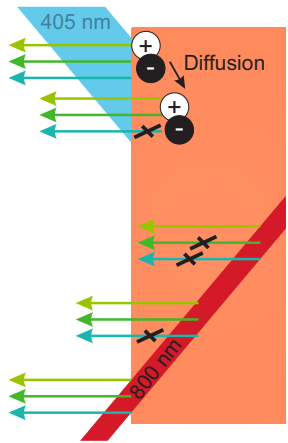}
\caption{Conventional PL and TPI-PL from (CH$_3$NH$_3$)PbBr$_3$. The different absorption lengths of 800~nm and 405~nm light provide a light source deep inside the crystal, and a reference spectrum from the surface-near region. Also indicated is diffusion of photoexcited carriers. 
}	 
\label{fig1}
\end{figure}

Perovskite single-crystals cleaved in ultrahigh vacuum (UHV) show  optical properties at room-temperature which are distinctly different from those of thin films and as-grown single crystals~\cite{murali2016}. Temperature-dependent measurements can help to develop a more complete picture of these optical properties,  which needs to include identification of exciton binding energies, free carriers, as well as direct and possible indirect transitions. Direct measurements of optical absorption spectra of single crystals in transmission are hindered by the short absorption length of visible light in OIPS~\cite{alias2016}, which forms a basis of their successful application in opto\-electronics. Measurements in reflectance are surface-sensitive, and easily affected by hydration of the surface~\cite{murali2016, fang2016}. To create clean surfaces and avoid their subsequent hydration, we perform experiments on (CH$_3$NH$_3$)PbBr$_3$ single crystals cleaved in UHV. 
Optical spectra of OIPS single crystals cleaved in UHV have, to the best of our knowledge, only been reported for (CH$_3$NH$_3$)PbBr$_3$ at room temperature~\cite{murali2016}. We develop an alternative approach to access  temperature-dependent optical spectra of OIPS. Therefore, steady-state PL spectroscopy at low excitation density and over a wide temperature range covering two phase transitions is performed on  surfaces of (CH$_3$NH$_3$)PbBr$_3$ single crystals cleaved in UHV. These surfaces emit broadband PL light. We also measure bulk-sensitive steady-state PL spectra induced by two-photon absorption (TPI-PL). Because of the vastly different excitation depths, indicated in Fig.~\ref{fig1}, absorption spectra can be extracted from the combination of PL and TPI-PL when absorption of TPI-PL light, as well as diffusion of carriers excited near the surface are taken into account. The procedure poses a viable route to extend temperature-dependent optical absorption spectroscopy to single crystals.

\begin{figure}
\center\includegraphics[width=0.7\columnwidth]{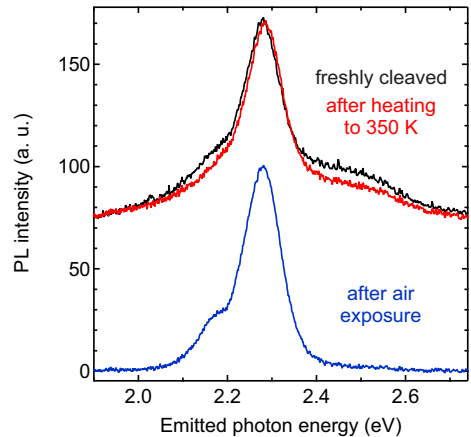}
\caption{PL spectra measured under different environmental conditions. 
}	 
\label{fig2}
\end{figure}

\section{Experimental details}

Details of crystal growth, characterization, and preparation are given in Ref.~\onlinecite{niesner2016}. Cleaving in UHV is used to create surfaces of (CH$_3$NH$_3$)PbBr$_3$ single crystals with bulk-terminated surface-Brillouin zone~\cite{komesu2016, niesner2016} and atomic arrangement~\cite{ohmann2015}. Crystals are cooled to 95~K after cleaving and kept at this temperature between measurements to avoid desorption of methylamine. 
The photoluminescence experiments are carried out at a base pressure $<10^{-7}$~Pa.  The experiments are illustrated in Fig.~\ref{fig1}.  For 405~nm excitation, a cw~laser source was used with an intensity  of 0.3~W/cm$^2$, comparable to the total terrestrial intensity of solar irradiation.  The PL spectra from  surfaces cleaved in UHV show distinct differences to the ones of air-exposed crystals and are much broader, see Fig.~\ref{fig2}. This broad emission spectrum forms the basis for further analysis. Data reported in this manuscript are reproducible over several heating and cooling cycles. Only keeping the crystals in UHV at room temperature for extended periods of time ($>10$~h) causes a narrowing of the spectra towards a spectral shape similar to the one measured from samples exposed to air.

The PL signal stimulated by 800~nm below band-gap excitation, Fig.~\ref{fig3}, is the result of two-photon absorption. 
We use a high repetition-rate laser (4.2~MHz) and low pulse energy (2~nJ) to ensure steady-state conditions, see Ref.~\onlinecite{suppl} for details. With the applied excitation density of $1\cdot 10^{13}$~cm$^{-3}$ absorbed photons per pulse the penetration depth (5~mm) of the laser is longer than the thickness of the crystals ($1.3\pm0.3$~mm). The excitation density is sufficient to saturate emission from defects that is most obvious in the low-temperature orthorhombic phase of OIPS. 

\section{Results and discussion}

\subsection{Excitation-density dependence and recombination channels}

\begin{figure}
\center\includegraphics[width=0.7\columnwidth]{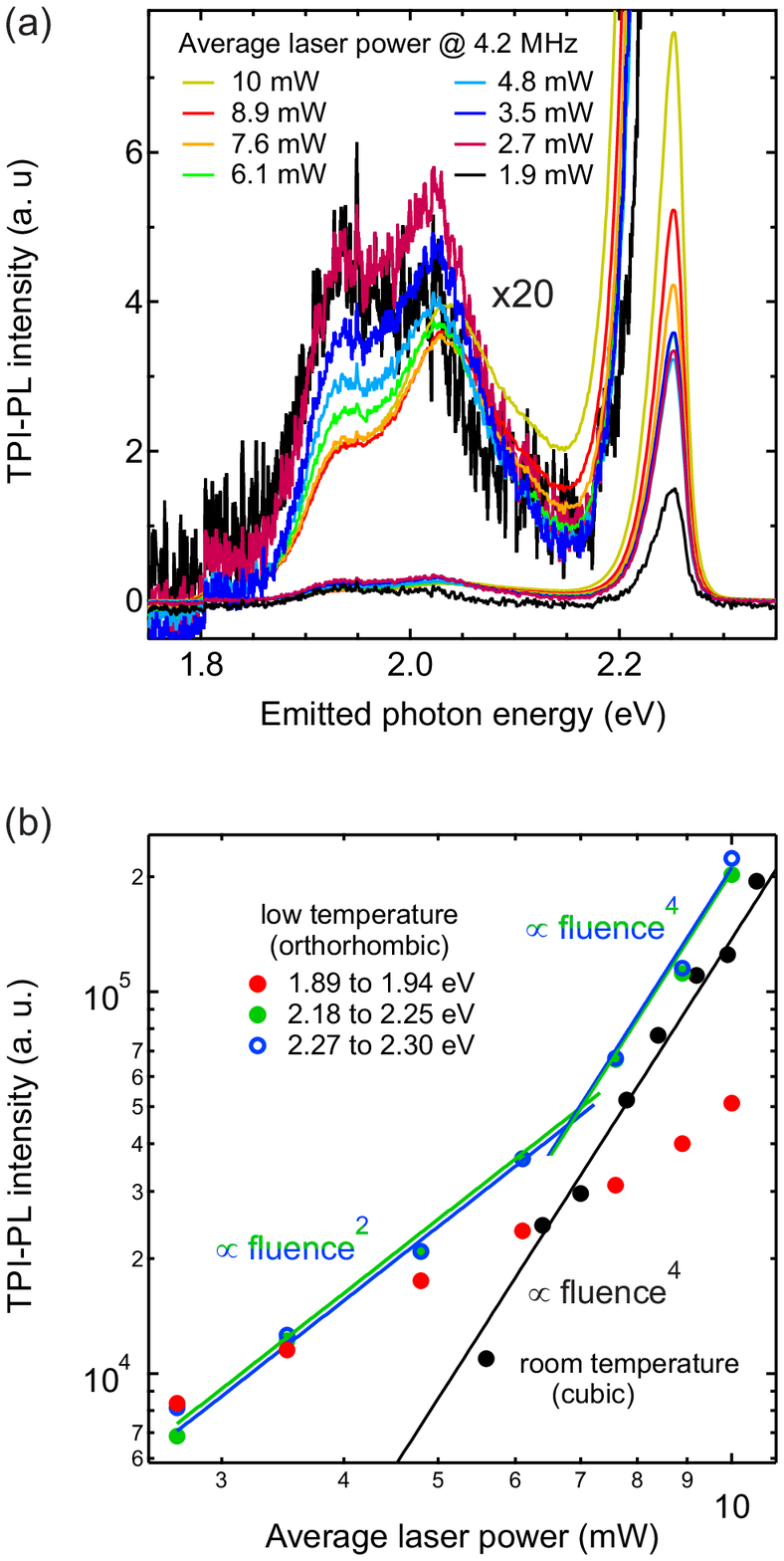}
\caption{Intensity-dependence of TPI-PL data. (a) shows spectra recorded on (CH$_3$NH$_3$)PbBr$_3$ in its low-temperature orthorhombic phase, normalized to the second power of the incoming laser flux. Intensities in selected spectral regions, as well as the integral TPI-PL intensity at room temperature are given in~(d). Red symbols give the intensity in the low-energy region. Green and blue symbols show the excitation density dependence of the main emission feature. Data for the room-temperature phase are given by black symbols.
}	 
\label{fig3}
\end{figure}

Figure~\ref{fig3}~(a) shows TPI-PL spectra of orthorhombic (CH$_3$NH$_3$)PbBr$_3$, normalized to the square of the laser fluence, which reflects the excitation density. The low energy ($<2.16$~eV) part of the spectrum saturates quickly, at a cw power of around 6~mW, see also Fig.~\ref{fig3}~(b). We thus attribute this part of the spectrum to defect emission. In this low excitation-density regime, the main two-photon induced photoluminescence emission signal grows quadratically with laser power, or linearly in excitation density, respectively.   
When the exciting laser power is increased to $>6$~mW, in contrast, the intensity of the main emission peak increases strongly, with the fourth power of the laser power, see Fig.~\ref{fig3}~(b). We ascribe the fourth-power dependence (quadratic in excitation density) to free carrier recombination~\cite{wang2016density}. As shown in Fig.~\ref{fig3}~(b), the quadratic increase of the TPI-PL intensity with excitation power is observed for all parts of the main TPI-PL emission peak, from its lowest energy region (2.18 to $2.25$~eV, green symbols) up to the highest emission energies (2.27 to $2.30$~eV, blue open symbols). We use the distinctly different excitation density dependences of the very low-energy part of the spectrum ($<2.16$~eV), which is characteristic for defects, and the main  part of the spectrum ($>2.16$~eV), characteristic for free carriers, to discriminate defect emission from intrinsic emission. 

Spectra shown in this paper were recorded with excitation densities beyond the onset of free-carrier emission, with defect emission saturated. The integral intensity of the room-temperature TPI-PL spectrum is given in Fig.~\ref{fig3}~(b) as well. The signal increases with the fourth power of laser fluence (quadratically in excitation density) down to the lowest TPI-PL intensities we can detect, with no indication of prominent photoluminescence emission from defect states. 

\begin{figure*}
\center\includegraphics[width=0.99\textwidth]{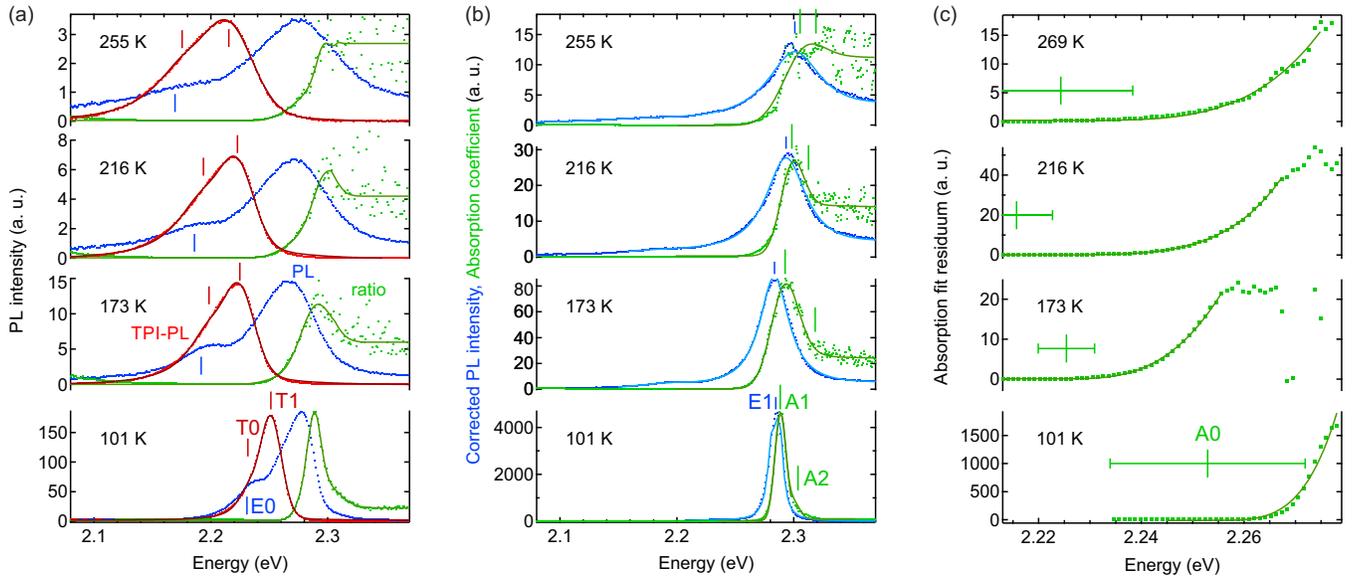}
\caption{(a) Photoluminescence spectra of single-crystal (CH$_3$NH$_3$)PbBr$_3$ after excitation with 405~nm light (blue) and 800~nm light (red), as well as their ratio (green). (b)~gives the extracted absorption coefficients and photoluminescence spectra corrected for reabsorption, together with fits to the absorption coefficients. Fit residua are given in~(c). Emission and absorption features described in the text are labeled. Their positions are marked by ticks. For the absorption onset A0 the error is indicated in addition.
}	 
\label{fig4}
\end{figure*}

\subsection{Extraction of absorption spectra}

In Fig.~\ref{fig4}~(a) we compare steady-state PL spectra with excitation wave\-lengths of 405~nm (blue) and 800~nm (red) for various temperatures. TPI-PL spectra are narrow with an asymmetric shape. The main emission occurs at lower energy than that of regular PL. We fit TPI-PL spectra using the sum of two Voigt functions T0 and T1 (dark red lines). Resulting peak positions are indicated by vertical lines. 
Two-photon absorption at wavelengths around 800~nm resembles one-photon absorption at half the wavelength~\cite{walters2015}, without being affected by real intermediate states or one-photon absorption. Therefore the PL emission following one-photon absorption at 405~nm and two-photon absorption at 800~nm should be almost identical for sufficiently thin samples, as observed experimentally for (CH$_3$NH$_3$)PbBr$_3$~\cite{gu2015} and CsPbBr$_3$~\cite{wang2015, xu2016} micro\-crystals. The main differences in case of extended crystals 
arise from the different penetration depths at 405~nm and 800~nm, followed by reabsorption of the PL light in the latter case, as illustrated in Fig.~\ref{fig1}. The detected PL emission $I(z, \lambda)$ from radiative recombination processes at a fixed distance $z$ from the surface is $I(z, \lambda)=I(0, \lambda) \exp\left(-\alpha(\lambda)\cdot z\right)$. Here $\alpha(\lambda)$ is the absorption coefficient. Both the absorption length for two-photon absorption and the thickness of the crystal are longer than the linear absorption length in the spectral range under investigation. We thus create and collect TPI-PL for $z=0$ to $\infty$, and the detected spectrum can be approximated by 

\begin{equation}\label{eq1}
I_{TPI-PL}(\lambda)=I(0, \lambda) \int_{0}^{\infty}\exp\left(-\alpha(\lambda)\cdot z\right) dz = \frac{I(0, \lambda)}{\alpha(\lambda)}
\end{equation}

Hence, if the PL spectrum at the surface $I(0, \lambda)$ is known, the absorption coefficient can be extracted from  TPI-PL. We use the PL spectrum $I_{PL}(\lambda)$ with 405~nm excitation as an estimate of $I(0, \lambda)$. The ratio $I_{PL}(\lambda)/I_{TPI-PL}(\lambda)$ is given by green dots in Fig.~\ref{fig4}~(a). The PL spectrum after 405~nm excitation, however, is itself subject to absorption~\cite{yang2015a, tian2015, yamada2016}, since carrier diffusion lengths in OIPS can exceed optical absorption lengths for photon energies above the band edge. Carriers hence recombine at finite distance from the surface, as illustrated in Fig.~\ref{fig1}. We account for the diffusion profile by integrating the PL intensity from different distances from the surfaces similar to Eq.~\ref{eq1}. The finite diffusion length $\sigma$ along the surface normal (the (001) direction of the crystal in the cubic phase) is modeled by introducing an additional term $\exp{\left(-\left(\frac{z}{\sigma}\right)^2\right)}$ in the integral. Lateral diffusion is neglected, since the spot size of the exciting laser ($> 3$~mm) is much larger than the carrier diffusion length~\cite{dong2015, shi2015, yamada2016}. The corrected PL spectrum $I_{PL}^{corr}$ is then given by 

\begin{equation}\label{eq2}
I_{PL}^{corr}=I_{PL}\left(\exp{\left(\left(\frac{\alpha\,\sigma}{2}\right)^2\right)}~\mathrm{erfc}\left(\frac{\alpha\,\sigma}{2}\right)\right)^{-1}. 
\end{equation}

We first use Eq.~\ref{eq2} to calculate the corrected PL spectrum from the measured PL spectrum (blue) and the ratio (green) shown in Figs.~\ref{fig4}~(a). The corrected PL spectra are given in Fig.~\ref{fig4}~(b) and for an extended energy range in Fig.~\ref{fig5}. From these corrected spectra and the TPI-PL data the corrected absorption coefficient is then calculated using Eq.~\ref{eq1}. Resulting absorption spectra are given in Figs.~\ref{fig4}~(b), (c). 

Correcting the absorption spectra for diffusion hardly changes the energetic positions of the absorption features discussed in the following, as illustrated in Ref.~\onlinecite{suppl}. Its only effect is a slight energetic shift of the absorption onset by a few meV towards higher energies, which is covered by the given error bars. Although the approximation $\alpha = {I_{PL}(\lambda)}/{I_{TPI-PL}(\lambda)}$  gives very similar results as the ones reported, we perform the correction for reasons of consistency. The influence of the correction for diffusion on the emission spectra is also illustrated in Ref.~\onlinecite{suppl}. The only peak position which is significantly altered as a result of diffusion is the one of the emission feature at 2.27~eV before the correction, or at 2.29~eV after, respectively.  
The peak position after correction matches the one found in PL from (CH$_3$NH$_3$)PbBr$_3$ micro\-crystals~\cite{gu2015, zhu2016}, in which diffusion and re-absorption play a minor role. 
While the exact line shape of the corrected spectra depends on the details of the model, the peak position at 2.29~eV is robust against changes in the diffusion profile, as long as $\alpha \sigma$ is sufficiently large. This is the case for  $\alpha \sigma\geq 3$ at 2.35~eV photon energy, as detailed in  Ref.~\onlinecite{suppl}. This requirement is fulfilled since $1/\alpha=90$~nm at 2.35~eV~\cite{yang2015c} and the diffusion length $\sigma$ in OIPS single crystals is in the $\mu$m range~\cite{dong2015, shi2015, yamada2016}. Corrected spectra in Fig.~\ref{fig4} are shown for $\alpha\,\sigma=5$ at 2.35~eV photon energy. $\alpha\,\sigma$ may be much longer because of photon recycling effects~\cite{pazos2016}, i.~e. multiple photon reabsorption and reemission. Remaining errors arising from the uncertainty in the exact diffusion profile are included in the error bars. They amount to 4~meV for the absorpion onset and are on the order of 1~meV for the other features, as detailed in Ref.~\onlinecite{suppl}.

\subsection{Modeling of absorption and emission spectra}

As proposed earlier~\cite{saba2014, yang2015c, yang2015a}, we model the absorption spectra $\alpha\left(E\right)$ using Elliot's theory~\cite{elliott1957, feneberg2013}: 

\begin{equation}
\begin{split}
	&\alpha\left(E\right)=
	 \frac{C_{A2}}{E^2} \frac{1+\mathrm{erf}\left(\left(E-E_{A2}\right)/\gamma\right)}{1-\exp\left(-2\pi^2\left|E_{x1}/\left(E-E_{A2}\right)\right|\right)}\\&+
	\frac{C_{A1}}{E^2}\sum_{n=1}^3 \left(\gamma n^3\right)^{-1} \exp\left(-\left(\frac{E-\left(E_{A2}-E_{x1}/n^2\right)}{\gamma}\right)^2\right)
\end{split}
\end{equation}

The first term describes band absorption A2 with an onset at $E_{A2}$ and amplitude $C_{A2}$, the second one excitonic enhancement A1 at $E_{A1}$ with amplitude $C_{A1}$. 
$\gamma$ is a broadening parameter, and $E_{x1}=E_{A2}-E_{A1}$ the exciton binding energy. Fits to the data are given as dark green lines in Fig.~\ref{fig4}~(b). 

The fits deviate from the absorption spectra systematically, as the measured spectra tail towards lower energy. Fit residua $\Delta \alpha(E)$ are given in Fig.~\ref{fig4}~(c). We fit a power law $\Delta \alpha(E) \propto \left(E-E_{A0}\right)^n ~\theta\left(E-E_{A0}\right)$ to the residua, where $E_{A0}$ is the position of the absorption onset A0 and $\theta\left(E\right)$ is Heavyside's theta function. The absorption onset lies energetically well above the emission from trap states identified from TPI-PL, shown in Fig.~\ref{fig3}. Fits give $n=4\pm 0.5$, consistent with an indirect transition in combination with a low density of states (DOS) at the band edges of OIPS found in 
calculations~\cite{jishi2014, umari2014, motta2015, menendez2014} and photo\-emission experiments~\cite{endres2016, niesner2016}. Power-law fits  are shown as dark green lines in Fig.~\ref{fig4}~(c), extracted ${A0}$ positions are marked by ticks.

\begin{figure}
\center\includegraphics[width=0.8\columnwidth]{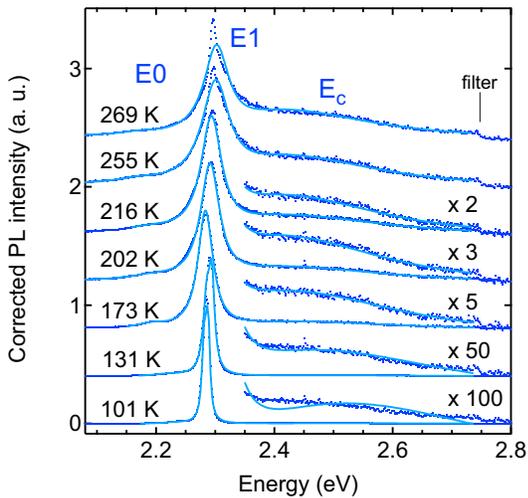}
\caption{Diffusion-corrected PL spectra and fits to the data. The emission peaks $E_0$ and $E_1$ are indicated. The intensity in the high-energy part of the spectrum is scaled for reasons of clarity. The scaling factors increase significantly in the low-temperature phase.
}	 
\label{fig5}
\end{figure}

\begin{figure}
\center\includegraphics[width=0.8\columnwidth]{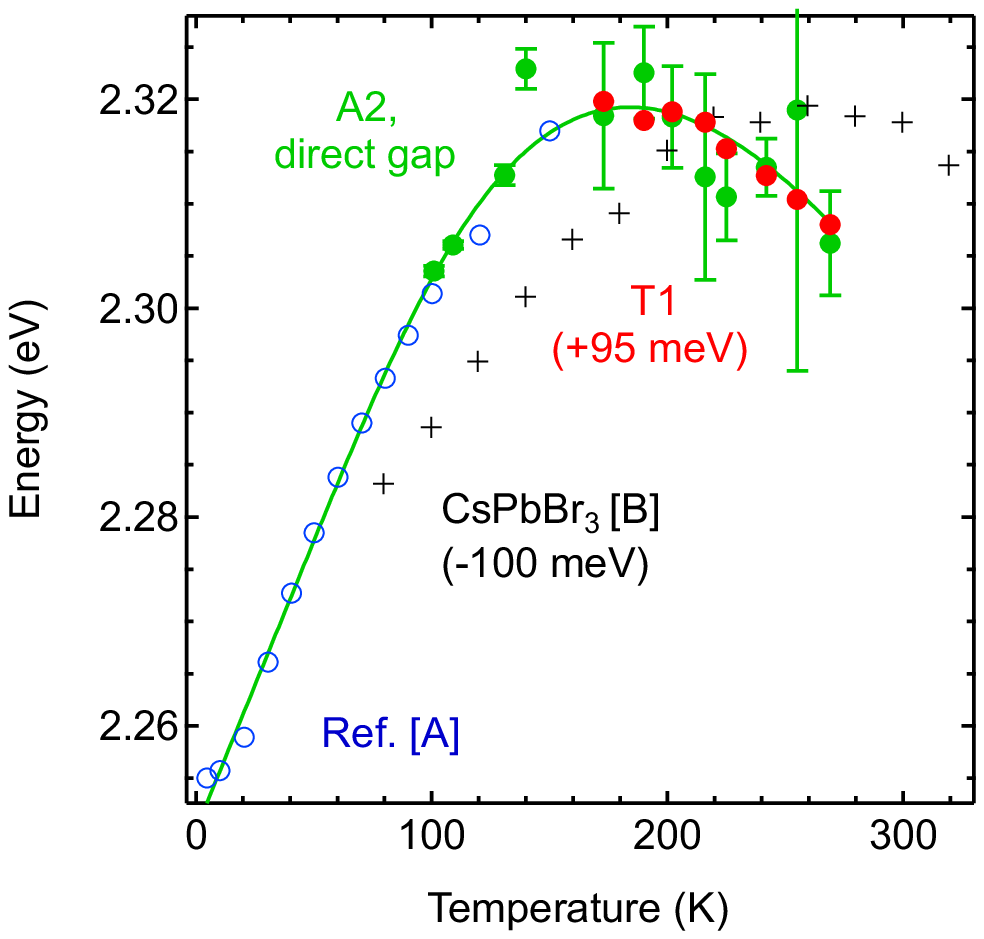}
\caption{Temperature-dependence of the direct band gap (green symbols). Data for the lower-energy transition in the tetragonal and cubic phases are also shown (red), and shifted in energy by 95~meV for better visibility. The green line represents a fit using the Bose-Einstein model. Data for the low-temperature phase of (CH$_3$NH$_3$)PbBr$_3$~\cite{tilchin2016}, as well as of related CsPbBr$_3$~\cite{wei2016}, shifted by $-0.1$~eV, are given for comparison. [A]: Ref.~\onlinecite{tilchin2016}, [B]:~Ref.~\onlinecite{wei2016}. 
}	 
\label{fig6}
\end{figure}

We now turn to the discussion of the PL spectra. As shown in Fig.~\ref{fig4} and~\ref{fig5}, the (corrected) PL spectra exhibit three structures: a prominent peak E1 at 2.29~eV, a low-energy peak E0, and a high-energy continuum E$_c$. We use Lorentzian functions to fit E0 and E1. We attribute the continuum emission to free carriers~\cite{zhu2016, niesner2016persistent}, and take them into account in the fit by adding the function~\cite{varshni1967}

\begin{equation}\label{eq4}
	f_c(E) dE \propto (E - E_{E1})^2 \exp\left(-E/kT^*\right) dE
\end{equation}
Where $E_{E1}$ is the energetic position of the E1 emission feature. $(E - E_{E1})^2$ reflects the joint density of states, which gives the emission spectrum after multi\-plication with the high-energy tail $\exp\left(-E/kT^*\right)$ of a Fermi-Dirac-like distribution. The parabolic density of states is introduced here empirically, since it gives a better fit to the data than the square-root function expected for a three-dimensional electron gas, as found also in photo\-emission experiments~\cite{endres2016, niesner2016}. 
Fits to the data are shown in Fig.~\ref{fig5}. The fitting range is constrained by trap emission below 2.16~eV and the cutoff of the long-pass filter at 2.75~eV. We give extracted fit parameters for free-carrier emission only for (CH$_3$NH$_3$)PbBr$_3$ in its tetragonal and cubic phase. Free carrier emission is weak at low temperature, in agreement with previous studies~\cite{zhu2016, niesner2016persistent}, see Fig.~\ref{fig5}. The  fitted exponential decay constant $kT^*\geq 0.06$~eV in Eq.~\ref{eq4} does not represent a thermodynamical temperature, but  reflects a non-equilibrium distribution of carriers. The average energy $\left\langle E_c\right\rangle$ of the carriers is shown in Fig.~\ref{fig6}~(a). For a parabolic density of states $\left\langle E_c\right\rangle = 3\,kT^*$, consistent with our data and ultrafast spectroscopies on (CH$_3$NH$_3$)PbI$_3$~\cite{niesner2016persistent}. 
For a square-root shaped DOS, in contrast, $\left\langle E_c\right\rangle = 1.5\,kT^*$. The low DOS at the band edges of OIPS provides one explanation for slow energy relaxation of carriers~\cite{kawai2015}. In time-correlated single photon counting from as-prepared (CH$_3$NH$_3$)PbBr$_3$ crystals, only a minor contribution of free carriers to the spectra was observed~\cite{zhu2016} as compared to their clear signature found here. The discrepancies probably arise  from different surface preparations. We find intense free-carrier PL only after cleaving in UHV. The feature disappears after exposure to air, see Fig.~\ref{fig2}. (CH$_3$NH$_3$)PbBr$_3$ is extremely sensitive to gas adsorption~\cite{murali2016, fang2016}, which can change surface band bending and trap state density, thus altering carrier diffusion and scattering. We point out the necessity to avoid exposure of OIPS to air and to prevent hydration of the surface prior to contacting, if energetic carriers are to be harvested in future experiments.

\subsection{Temperature-dependence of the direct band gap}

Our analysis determines the temperature dependence of the direct band gap of cleaved (CH$_3$NH$_3$)PbBr$_3$ single crystals, which is given in Fig.~\ref{fig6}. The band gap increases slightly with temperature below 200~K, and decreases at higher temperatures. This behavior has been observed for other semiconductors~\cite{bhosale2012} including CsPbBr$_3$~\cite{wei2016}, and can be understood within the Bose-Einstein model~\cite{lautenschlager1987}:

\begin{equation}
\begin{split}
 &E_{A2}(T) - E_{A2}^{0}=\frac{\partial E_{A2}(T)}{\partial V} \frac{\partial V}{\partial T}T\\&+A_{EP}\left( \frac{2}{\left(\exp\left(\hbar\omega / kT \right)-1\right)} +1 \right)
\end{split}
\end{equation}
The first term in the sum describes the change in the band gap associated with the thermal expansion of the lattice. $\frac{\partial V}{\partial T}$  of (CH$_3$NH$_3$)PbBr$_3$ is positive with a relative value $\frac{1}{V} \frac{\partial V}{\partial T} = 1.0\cdot 10^{-4}$~K$^{-1}$ for temperatures $>100$~K~\cite{Swainson2003}. Calculations show that the band gap increases with increasing lattice constant~\cite{dar2016, meloni2016}, making the slope of the linear term overall positive. The second term in the sum accounts for the renormalization of the band gap by electron-phonon coupling. $A_{EP}$ is the strength of the coupling and $\omega$ the frequency of the prominently coupling phonon mode. $A_{EP}$ is typically negative and gives rise to a decrease in band gap at elevated temperature~\cite{bhosale2012, wei2016}, as also observed here for cleaved  (CH$_3$NH$_3$)PbBr$_3$ single crystals. A fit to the data is given by the green solid line in Fig.~\ref{fig6}. The resulting parameters are $E_{A2}^{0}=2.55\pm 0.07$~eV, $\frac{\partial E_{A2}(T)}{\partial V} \frac{\partial V}{\partial T} = (5.5\pm 0.4) \cdot 10^{-4}$~eV/K, $A_{EP} = -0.30 \pm 0.07$~eV and $\hbar\omega = 47 \pm 6$~meV. The phonon energy matches the one of the 40~meV optical mode observed in Raman spectroscopy~\cite{yaffe2016, letoublon2016}. We note that our measurements span several structural phases of (CH$_3$NH$_3$)PbBr$_3$. The direct band gap does not seem to change across those phase transitions. 

In addition to our data and the Bose-Einstein model, Fig.~\ref{fig6} gives the temperature-dependent band gap of CsPbBr$_3$~\cite{wei2016} for comparison, as well as the direct band gap determined for (CH$_3$NH$_3$)PbBr$_3$ single crystals at low temperature from photoluminescence spectroscopy at low excitation density~\cite{tilchin2016}. In the latter experiment the band gap was identified by resolving two excitonic transitions in the low-temperature spectra. Our analysis shows excellent agreement with these data in the temperature range which is covered by both experiments. Extra\-polation of the fit of the Bose-Einstein model to our data gives good agreement with the reported low-temperature data as well. On the contrary, our observation of a decreasing band gap at elevated temperature is in apparent contrast to PL measurements performed at high excitation density on (CH$_3$NH$_3$)PbBr$_3$ polycrystalline films~\cite{wright2016, dar2016}, which observe a monotonous blue-shift of the main emission peak with increasing temperature. Whether this discrepancy is the result of different electron-phonon coupling in cleaved single crystals as compared to thin films or the result of band filling effects related to the excitation density~\cite{dar2016} is to be investigated.

\begin{figure*}
\center\includegraphics[width=0.99\textwidth]{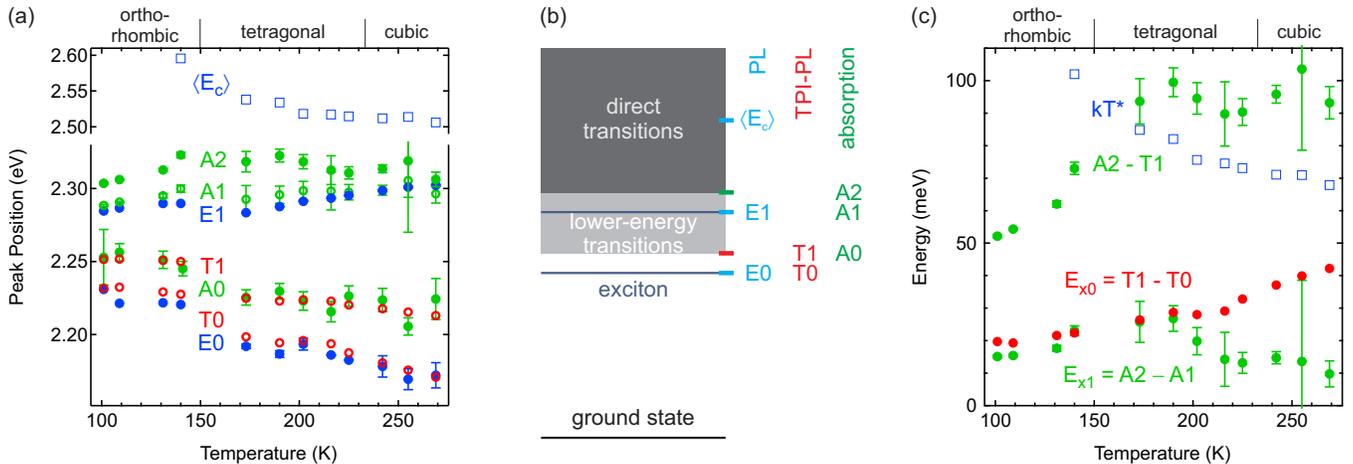}
\caption{Results of fits to PL emission (blue: E0, E1, E$_c$), TPI-PL data (red: T0, T1) and extracted absorption spectra (green: A0, A1, A2). (a) gives the extracted peak positions as discussed in the text and as labeled in Figs.~\ref{fig4} and~\ref{fig5}. A schematic energy level diagram is given in (b).  
In (c) spacings between selected peaks are given  by symbols. (c) also shows the carrier quasi-temperature (open squares). 
}	 
\label{fig7}
\end{figure*}

\subsection{Summary of extracted transition energies}

The experimentally determined positions of all spectral features are summarized in Fig.~\ref{fig7}~(a). A schematic energy level diagram is given in Fig.~\ref{fig7}~(b). PL emission peaks are indicated in blue (E0, E1, and E$_c$), peak positions from TPI-PL in red (T0 and T1), absorption features are given in green (A0, A1, and A2). The prominent excitonic transition A1 (E1) occurs at 2.29~eV. This value matches the one reported from PL experiments on (CH$_3$NH$_3$)PbBr$_3$ micro\-crystals~\cite{gu2015, zhu2016}, in which diffusion and re-absorption play a minor role. The agreement gives experimental evidence for the validity of our data analysis. Moreover, the related exciton binding energy $E_{x1}=E_{A2}-E_{A1}=22\pm 2$~meV for $T\leq 200$~K, Fig.~\ref{fig7}~(c), is in agreement with the value of $25\pm 5$~meV reported from low-temperature magneto\-absorption~\cite{galkowski2016}.

While the position of the absorption onset A0 in Fig.~\ref{fig7}~(a) follows a similar temperature dependence as the A2 direct transition within the individual phases, it changes at the orthorhombic-to-tetragonal phase transition from 2.25~eV to around 2.22~eV. Emission from the transition associated with the A0 absorption onset is observed in TPI-PL (T1). Its position at room temperature matches the absorption onset found in transmission from the same crystals~\cite{niesner2016}, the band gap found in reflectance measurements on single-crystals cleaved in UHV~\cite{murali2016}, and the onset reported from photo\-luminescence excitation~\cite{fang2016}. The transition is accompanied by excitonic emission observed both in regular PL (E0) and TPI-PL (T0).  
The temperature-dependent exciton binding energy $E_{x0}=E_{T1}-E_{T0}$ is given in Fig.~\ref{fig7}~(c). It matches $E_{x1}$ for $T\leq 200$~K. At higher temperature $E_{x1}$ decreases, as observed also for (CH$_3$NH$_3$)PbI$_3$~\cite{even2014a, yamada2015, miyata2015, fang2015a, even2016}. In contrast, E0 and T0 redshift, possibly due to a reduction of effective mass~\cite{azarhoosh2016} and enhanced vibronic coupling, i.~e. \mbox{(multi-)}phonon emission in an indirect emission process and large polaron formation~\cite{ma2014, bokdam2015, Menendez2015, Frost2016, zhu2016}.

\subsection{Discussion of the two observed transitions}

The question arises, why the prominent direct transition is found several tens of meV above the absorption onset. The exact energy difference is given as A2$-$T1 in Fig.~\ref{fig7}. PL spectra similar to the ones presented here, with two emission peaks and a signature of free carriers, were measured on single-crystal (CH$_3$NH$_3$)PbI$_3$ at low temperature~\cite{diab2016}. The authors assign the lower-energy emission peak to defects. In case of (CH$_3$NH$_3$)PbBr$_3$, it is unlikely that the lower-energy transition arises from defects, for a number of reasons. The transition lies energetically well above the defect emission which is identified at energies $\leq 2.16$~eV  from excitation-density dependent TPI-PL data, Fig.~\ref{fig3}. Exposure of the samples to air and extended heating in UHV result in a blue shift of the low-energy emission feature and an increase of the band gap, see Fig.~\ref{fig1}, as also storing the samples in UHV at room temperature for extended periods of time ($>10$~h) does. Heating (CH$_3$NH$_3$)PbBr$_3$ in UHV to $>320$~K results in desorption of methylamine~\cite{niesner2016}, and can be expected to increase the density of defect states rather than reducing it.  Moreover, photocurrent measurements showed that 2.18~eV excitation (matching the E0 and T0 energies) creates mobile charges in (CH$_3$NH$_3$)PbBr$_3$ at room temperature~\cite{fang2015}. This process would be inefficient if optically active defect states were excited initially, and then needed to be thermally activated to the bands related to the transitions around 2.3~eV with an activation energy of $\approx 120$~meV. In a previous study~\cite{wu2016}, as-grown surfaces of single-crystal (CH$_3$NH$_3$)PbBr$_3$ were also investigated using a combination of TPI-PL and conventional PL. The authors assign low (high) energy emission to the bulk (surface) of the crystal. 
This interpretation is supported by a recent study~\cite{murali2016}, that systematically compares as-grown crystals with the same crystals after cleaving in UHV. The authors find a band gap of 2.22~eV of cleaved surfaces as opposed to 2.27~eV for as-grown crystals. The absorption onset at 2.22~eV in our room-temperature measurements is in excellent agreement with the former value, giving further  evidence that a band gap of 2.22~eV is indeed an intrinsic property of  (CH$_3$NH$_3$)PbBr$_3$ single crystals cleaved in UHV. 

We note that the absence of the higher-energy E1 and E$_c$ features in the TPI-PL data does not exclude that recombination of high-energy excitons and carriers takes place deep in the crystal, and that their luminescence signals get absorbed before they reach the surface. Indeed, the TPI-PL intensity reaches zero only for energies around 2.4~eV, indicating free-carrier recombination in the bulk.

A possible explanation for the coexistence of two optical transitions in the near band-edge optical spectra of (CH$_3$NH$_3$)PbBr$_3$ could be lateral inhomogeneities and the coexistence of several structural phases at the surface after cleaving. Phase coexistence at temperatures below the one of the orthorhombic-tetragonal phase transition has been observed experimentally both using bulk-sensitive~\cite{galkowski2016a, dar2016} and surface-sensitive~\cite{ohmann2015} techniques. While coexisting structural phases at the surface seem less likely in the tetragonal and cubic phases, they can not be ruled out on the basis of our data. 

The coexistence of a direct and an indirect band gap has also been observed in (CH$_3$NH$_3$)PbI$_3$, with energetic spacings of 47~meV~\cite{hutter2016} and 60~meV~\cite{wang2016}. 
The observation of direct and indirect gaps for both (CH$_3$NH$_3$)PbI$_3$ and (CH$_3$NH$_3$)PbBr$_3$ indicates that they might be a common property of both OIPS, originating from their similar band structures~\cite{mosconi2016}. Calculations find a slightly indirect band gap for both OIPS~\cite{even2013,kim2014, zheng2015,etienne2016,azarhoosh2016, mosconi2016} as a consequence of Rashba-type spin-split bands. Rashba splitting causes a shift of spin-polarized electronic bands in k-space. Different spin splittings in the valence and conductance band result in a mismatch of the states at the band edges in momentum and spin-polarization, no longer allowing for direct optical transitions. The situation is illustrated schematically in Ref.~\onlinecite{suppl}. As Rashba splitting has been observed experimentally for (CH$_3$NH$_3$)PbBr$_3$~\cite{niesner2016} and proposed as the origin of an indirect gap in (CH$_3$NH$_3$)PbI$_3$~\cite{wang2016}, the coexistence of direct and indirect transitions poses another possible explanation for the measured spectra.  
\section{Summary}

In summary, we combine bulk-sensitive two-photon induced photoluminescence and more surface-sensitive conventional photoluminescence from  (CH$_3$NH$_3$)PbBr$_3$ single crystals cleaved in ultrahigh vacuum. The surfaces show broadband emission. In combination with the strongly different information depths of one- and two-photon induced photoluminescence, the broad PL spectrum allows us to extract absorption spectra. The technique is well compatible with common ultrahigh vacuum apparatus. Extracted absorption spectra reveal a prominent direct transition at 2.31~eV, as well as an additional transition at 2.25~eV (2.22~eV) in the orthorhombic (tetragonal) phase, each accompanied by excitonic emission. Defect emission is identified at photon energies $<2.16$~eV. 
We perform temperature-dependent measurements, and determine the optical band gap and exciton binding energies as a function of temperature across several phase transitions. The temperature-dependence of the band gap is well described by a Bose-Einstein model that takes into account band gap renormalization by electron-phonon coupling to vibration with a phonon energy of 47~meV. While excitons are found 22~meV below both band transitions at low temperature, they exhibit opposite temperature dependences at elevated temperature $> 200$~K. The findings are consistent with a slightly indirect band gap, which has been proposed as a source of long-lived carriers in OIPS~\cite{zheng2015, etienne2016, azarhoosh2016}. In addition, high-energy emission from free carriers centered around 2.5~eV is found from freshly cleaved single crystals. The feature disappears after exposing the crystals to air. The results underline the importance of environmental conditions for optical spectroscopy on OIPS and during the fabrication of contacts in devices. 

\bibliography{perovskites}

\end{document}